\documentclass[%
 preprint,
 amsmath,amssymb,
 aps,
 prb,
]{revtex4}

\usepackage{graphicx}% Include figure files
\usepackage{dcolumn}% Align table columns on decimal point
\usepackage{bm}% bold math
\usepackage{graphicx}
\usepackage{float}
\usepackage{natbib}
\usepackage{subfigure}
\usepackage{color}
\usepackage[mathlines]{lineno}% Enable numbering of text and display math
\usepackage{verbatim}

\begin{document}

%\preprint{APS/123-QED}

\title{Lifetime and Polarization for Real and Virtual Correlated Stokes-anti-Stokes Raman Scattering in Diamond}% Force line breaks with \\

\author{Filomeno S. de Aguiar J\'unior$^1$}
 \email{filomeno@fisica.ufmg.br}
\author{Marcelo F. Santos$^2$}
\author{Carlos H. Monken$^1$}
\author{Ado Jorio$^1$}%%
\email{adojorio@fisica.ufmg.br}
\affiliation{$^1$Departamento de F\'{\i}sica, ICEx, Universidade Federal de Minas Gerais, Avenida Antonio Carlos, 6627, Belo Horizonte, Minas Gerais 31270-901, Brazil}%

 \affiliation{$^2$Instituto de F\'{\i}sica, UFRJ, CP 68528, Rio de Janeiro, Rio de Janeiro 21941-972, Brazil}%

\date{\today}% It is always \today, today,

\begin{abstract}
The production of correlated Stokes (S) and anti-Stokes (aS) photons (SaS process) mediated by real or virtual phonon exchange has been reported in many transparent materials. In this work, we investigate the polarization and time correlations of SaS photon pairs produced in a diamond sample. We demonstrate that both S and aS photons have mainly the same polarization of the excitation laser. We also perform a pump-and-probe experiment to measure the decay rate of the SaS pair production, evidencing the fundamental difference between the real and virtual (phonon exchange) processes. In real processes, the rate of SaS pair production is governed by the phonon lifetime of $(2.8\pm 0.3)\ \mathrm{ps}$, while virtual processes only take place within the time width of the pump laser pulses of approximately $0.2\ \mathrm{ps}$. We explain the difference between real and virtual SaS processes by a phenomenological model, based on probabilities of phonon creation and decay.
\end{abstract}

\maketitle

%\tableofcontents

\section{Introduction}
\label{s:intro}

In correlated Raman scattering, the same phonon participates in both Stokes (S) and anti-Stokes (aS) frequency conversions, characterizing the SaS process~\cite{Klyshko1977a,Parra-Murillo2016}. The phonon created in the Stokes process is annihilated by the anti-Stokes one, generating a time-correlated photon pair. The phonon generated in a resonant S scattering has a lifetime $\tau_{p}$, which is typically of the order of a few picoseconds, and the correlated photon pair is created only if the aS scattering happens within a delay time not much longer than $\tau_{P}$ \cite{Lee2011e}. This so-called \textit{real} SaS scattering has been studied in several materials \cite{Jorio2014, Kasperczyk2015, Kasperczyk2016},  explored as the implementation of a Raman quantum memory for light in diamond \cite{Reim2010a, Lee2011e, England2013a} and gases \cite{Reim2011, VanderWal2003, Murphy1977}, and used to measure the lifetime of one-phonon Fock state, of the order to 3.9-ps \cite{Anderson2018a,PhysRevX}.  
	
In recent studies, it was shown that the formation of SaS correlated photon pairs can also occur mediated by the exchange of virtual phonons, which is referred to as \textit{virtual} SaS processes~\cite{Saraiva2017}. The photon pair produced by a virtual process is analogous to the electronic Cooper pair in superconductivity \cite{BCS}, and this analogy has been explored in diamond samples \cite{Junior2019, DeAguiarJunior2019}. The \textit{virtual} SaS emerges as a source of correlated photon pairs in a wider range of energies, different from the \textit{real} SaS that is restricted to $E_L\pm E_{P}$, where  $E_L$ is the excitation laser energy and $E_{P}$ is the phonon energy.

While the \textit{real} SaS process has a characteristic time scale dictated by $\tau_{P}$, in the \textit{virtual} SaS process the exchange of virtual phonons is expected to be nearly instantaneous, limited by the inverse bandwidth of the excitation pulse. In this work, we study the production rate of photonic Cooper pairs as a function of the time delay between the S and aS scattering in the SaS process to elucidate this fundamental difference between the real and virtual phonon exchange processes. As it will be discussed, photon polarization has also to be studied to enable performing the lifetime measurements.
\begin{figure}[]
	\centering
	\includegraphics[scale=0.29]{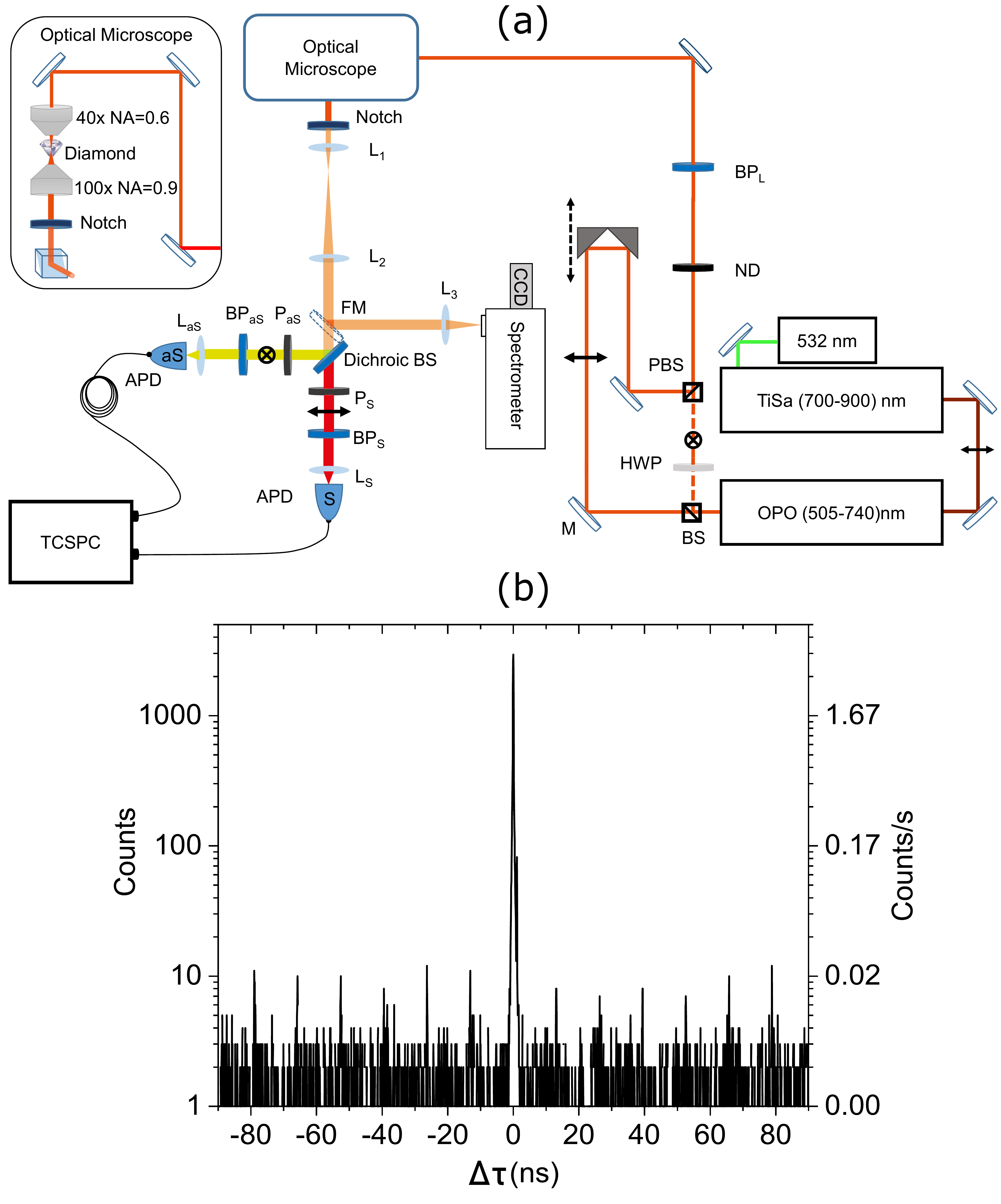}
	\caption{(a) Representation of the experimental setup. M: mirror; BS:  50:50 beam splitter; HWP: half-wave plate; PBS: polarization beam splitter; ND: neutral density filter;  BP: interference band-pass filter; L: lens; FM: flip mirror;  P: linear polarizers; TCSPC: time-correlated photon counting system. The subscript L stands for laser, S for Stokes and aS for anti-Stokes. The double-arrow dashed black line indicates the displacement of the mirrors to generate the pump-probe delay. The double-arrow solid lines and the crossed-circles indicate the polarizations of the excitation laser and scattered light. (b) Histogram of time delay $\Delta\tau$ between the arrival of S and aS photons in diamond at $\pm 900\ \mathrm{cm}^{-1}$ for a total of 600 seconds accumulation, with $16.6\ \mathrm{mW}$ laser excitation power at the sample. Each histogram bin has a time width of $96\ \mathrm{ps}$. This histogram was acquired in the $(H;H,H)$ configuration (see text). \label{fi}}
\end{figure}

\section{Technical details}
\label{s:tech}

Figure \ref{fi}(a) shows the experimental setup. A pulsed Ti:Sapphire laser, with repetition rate of $76\ \mathrm{MHz}$, excited by  a $10\ \mathrm{W}$ at $532\ \mathrm{nm}$ laser, produces a train of pulses of approximately $200\ \mathrm{fs}$ nominal width. The Ti:Sapphire laser is fixed at $800\ \mathrm{nm}$ and it is used as pump to an OPO (Optical Parametric Oscillator), which converts the $800\ \mathrm{nm}$ laser pulses into $633\ \mathrm{nm}$ pulses with intensity profile $w_L$ of approximately $0.4\ \mathrm{ps}$.

For the lifetime measurements, the $633\ \mathrm{nm}$ pulse train is divided in two by a  50/50 beam-splitter(BS), both with  polarization horizontal (H) with respect to the optical table. One of these pulsed beams passes through a delay line and has its polarization changed to vertical (V) by a half-wave plate (HWP). The two pulse trains are then recombined into a single beam by a polarization beam splitter (PBS) and directed to an inverted microscope, where the sample is located. The combined beam crosses neutral density filters (ND) used to reduce the power, and a band-pass filter (BP$_\mathrm{L}$) used as a laser line filter for $633\ \mathrm{nm}$. 

An objective of 40$\times$ magnification and numerical aperture NA = 0.60 mounted on top of the inverted microscope (see inset to Fig. 1(a)) focuses the combined beam in a diamond sample (Type IIac – Diacell design–(100)-oriented, https://www.almax-easylab.com/TypeIIacDiacellDesign.aspx). In the forward scattering geometry, another objective of  100$\times$ magnification and NA = 0.9 collects the forward scattered light. Notch filters are used to block the Rayleigh scattering. The Raman signal may be directed to a spectrometer by a flipping mirror FM, to analyze the emission spectrum of the sample. Once the flipping mirror is switched, the Raman signals go to a dichroic beam splitter, which reflects the anti-Stokes (aS) component to the aS APD (avalanche photodiode) and transmits the Stokes (S) component  to the S APD. The signals from the two APDs are sent to a time correlator  (PicoHarp 300), which records the detection time of the pulses from both APDs and builds a histogram of the time delay between the Stokes and anti-Stokes pulses in bins of $192\ \mathrm{ps}$. This time binning is compatible with the resolving time of the APDs.

The evidence of occurrence of correlated SaS processes is the presence of a distinguished peak at the central bin ($\Delta \tau =0\pm 96\ \mathrm{ps}$) of the histogram (Figure 1(b)) with total counting number more than twice as high as the set of the other peaks. The coincidence counts corresponding to S and aS photons scattered by the same phonon will fall within this central peak \cite{Parra-Murillo2016,Lee2011e,Jorio2014, Kasperczyk2015, Kasperczyk2016, Reim2010a,England2013a, Reim2011}, referred to as $\Delta\tau=0$ hereafter.
 
\section{Results}
\label{s:res} 
 
\subsection{Polarization}
\label{s:resPol}

To study the polarization dependence of the SaS process, we chose to excite the sample with H polarized light pulses. The polarization of the S and aS photons are then investigated using two linear polarizers ($\mathrm{P_S}$ and $\mathrm{P_{aS}}$), placed in front of each APD (see Fig.\ref{fi}(a)). Considering the representation ($L_{Pol.}$; $S_{Pol.}$, $aS_{Pol.}$) for laser, Stokes and anti-Stokes polarizations, we measured coincidence counts for the $(H; H, H)$, $(H; V, H)$, $(H; H, V)$, and $(H; V, V)$ scattering geometries.

\begin{figure}
	\centering
	\includegraphics[scale=0.7]{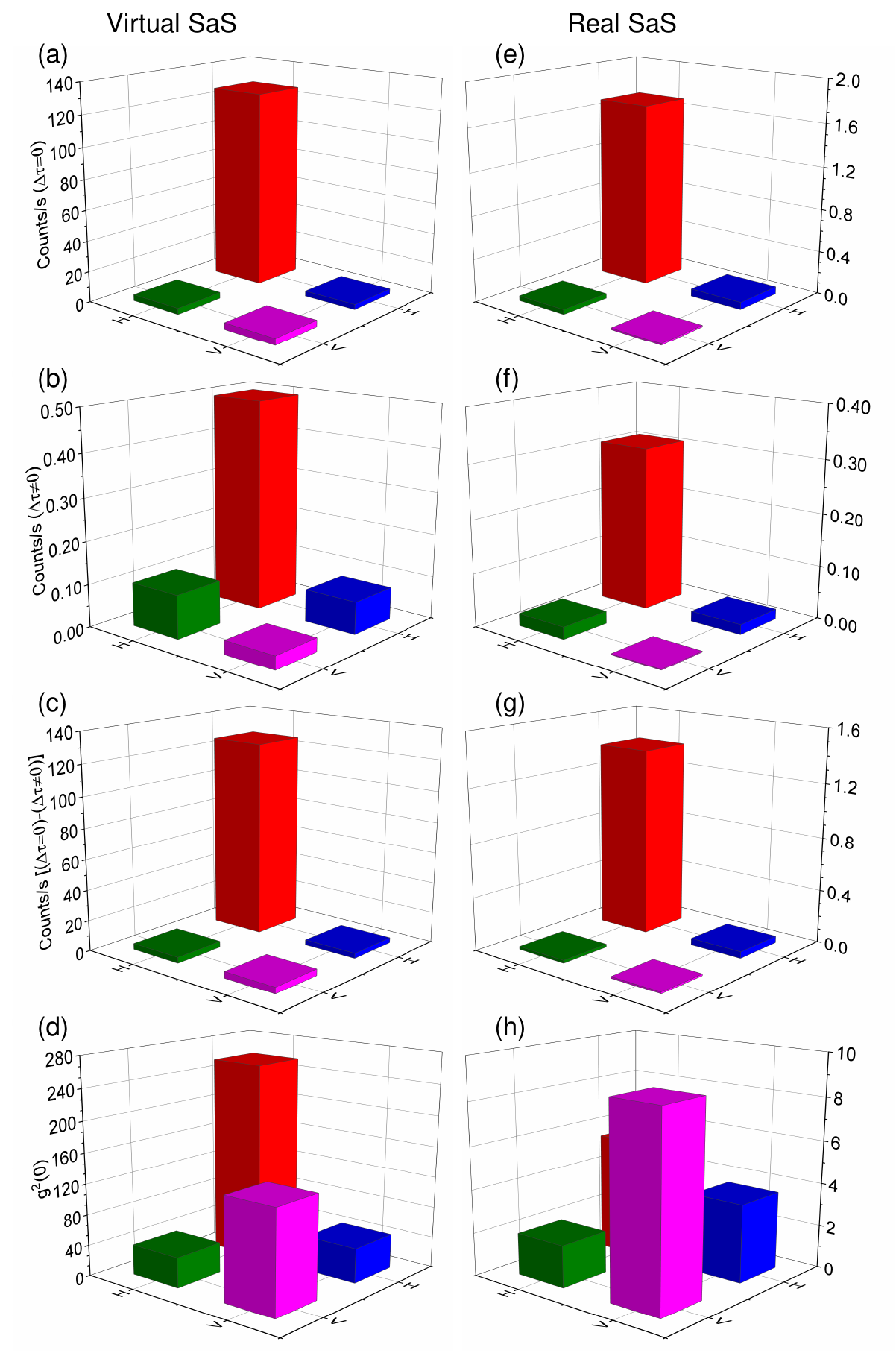}
	\caption{\label{fig1} Polarization analysis for the real SaS process at $\pm(1348 \pm 57)\ \mathrm{cm}^{-1}$  (right panels),  with $3\ \mathrm{mW}$ at the sample, and for the virtual SaS process (left panels), with Raman shift at $ \pm(900 \pm 133)\ \mathrm{cm}^{-1} $, with $42\ \mathrm{mW}$ at the sample. An iris was used to filter the accidental coincidences in the real process due to uncorrelated Raman scattering at higher angles \cite{Junior2019}. (a,e) Production rate obtained by the number of counts at $\Delta\tau =0 $, divided by the total detection time ($900\ \mathrm{s}$ for virtual and $300\ \mathrm{s}$ for real). Each bar corresponds to a measurement with a different aS and S polarization. In (b,f) we show the uncorrelated counts, calculated by the average of the peaks at $\Delta\tau \neq 0$, normalized by the detection time. (c,g) Correlated SaS, calculated by subtracting the averaged uncorrelated counts ($\Delta\tau \neq 0 $) from the counts at $\Delta\tau =0 $. (d,h) Cross correlation function $g^{(2)}_{S,aS}(0)$, calculated by the ratio between the counts at $\Delta\tau =0 $ and the average of uncorrelated counts at $\Delta\tau \neq 0 $.}\label{fig1a}
\end{figure}

We investigated the polarization in the SaS emission for both the \textit{real} process, when the Stokes and anti-Stokes energies correspond to the first-order Raman peak from diamond at $\pm 1332\ \mathrm{cm}^{-1}$ (Fig.\ref{fig1a}, right panels), and for the \textit{virtual} process, placing the S and aS band pass filters at $\pm 900\ \mathrm{cm}^{-1}$ (Fig. \ref{fig1a}, left panels). We observe that the SaS emission occurs majorly in the same polarization of the laser $(H; H, H)$, with the rate of SaS production (counts per second at $\Delta\tau =0, $ see panels (a) to (e)) being much higher in comparison with the other configurations for both \textit{virtual} and \textit{real} processes. In approximately 93\% (91.5\%) of the pairs in the \textit{real} (\textit{virtual}) SaS process, both Stokes and anti-Stokes photons have the same polarization of the laser. The creation of pairs $(H; V, V)$ with polarization orthogonal to the laser is less than 3\% of the total. Similar rates are obtained for pairs where the Stokes and anti-Stokes photons have orthogonal polarization with respect to each other, $(H; H, V)$ and $(H; V, H)$. For the uncorrelated process ($\Delta\tau \neq 0$, panels (b) and (f)), the maximum counts also happen for polarizations $(H; H, H)$, and the minimum for $(H; V, V)$. The accidental counts increase when the Stokes photons, or the anti-Stokes photons, or both have the same polarization of the laser. For completeness,  Fig. \ref{fig1a} (d,h) show the measured second-order cross-correlation function $g_{SaS}^{(2)}(\Delta\tau=0)$, which is $g_{SaS}^{(2)}(\tau)$ integrated in the central ($\Delta\tau=0$) histogram peak.

\subsection{Lifetime}
\label{s:resLif} 

Based on the polarization results, the lifetime of the phonons participating in the SaS process can be investigated analyzing the SaS delayed coincidence count rates as a function of the time delay between cross-polarized excitation pulses. In a pump-probe configuration, a first laser pulse (H-Pulse), horizontally polarized, generates a Stokes scattered photon, creating a phonon in the sample; a photon from a vertically polarized second laser pulse (V-Pulse), which can be delayed with respect to the H-pulse by a delay line, annihilates this phonon, generating the anti-Stokes scattered photon, creating the SaS pair. Considering that in the SaS process the Stokes and anti-Stokes scattered photons have predominantly the same polarization of the excitation laser (Fig. \ref{fig1} panels (a) and (e)), the S photons from the first pulse are majorly H-polarized, and directed to the S APD, while the aS photons created in the second pulse are majorly  V-polarized and directed to the aS APD. 

\begin{figure}[]
	\centering
	\includegraphics[scale=0.37]{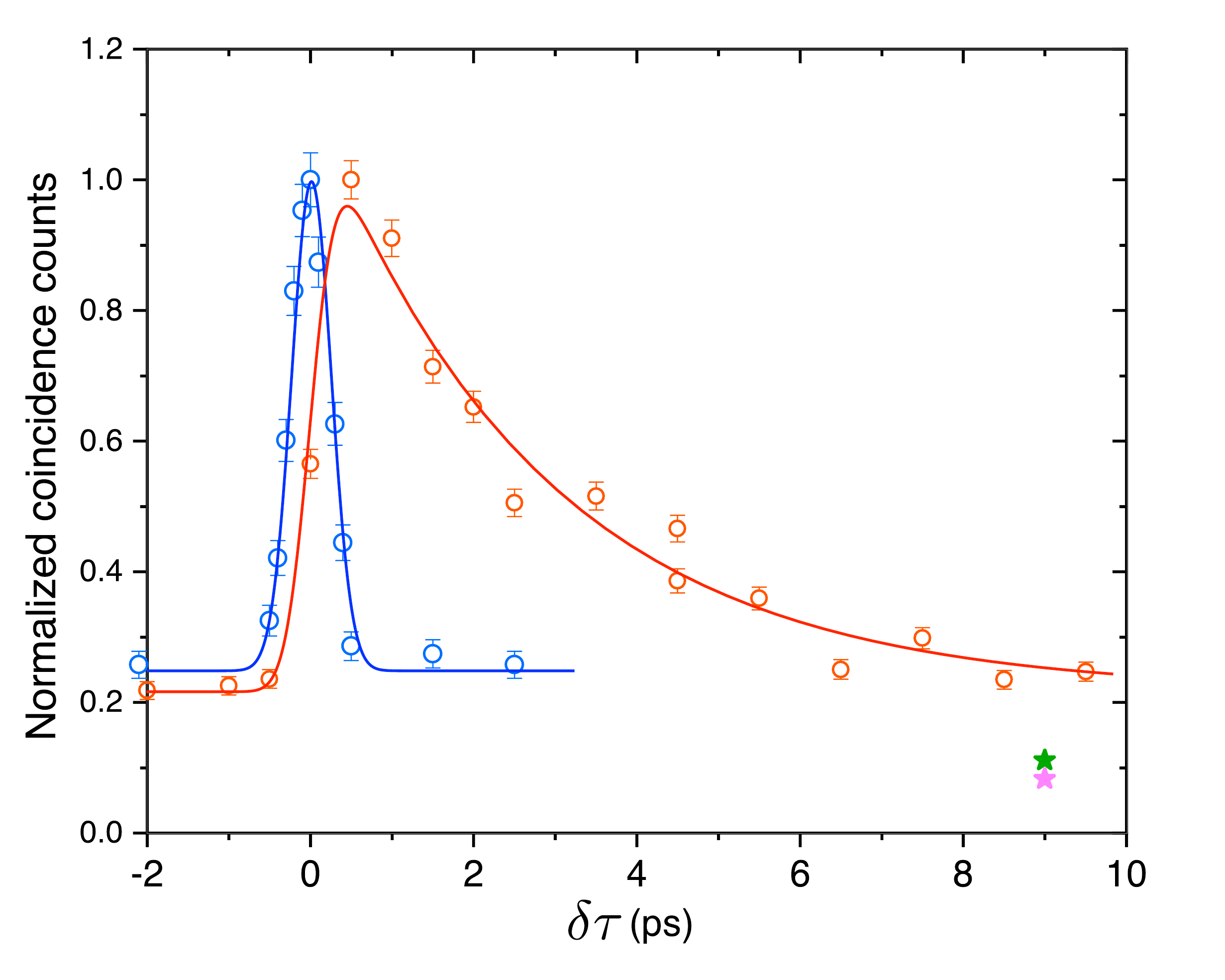}
	\caption{SaS scattering in diamond as a function of the time delay $\delta\tau$ between the H and V laser pulses. The blue circles stand for the SaS intensities of the virtual process at $\pm 900\ \mathrm{cm}^{-1}$, and the red circles are the SaS intensities for the real process at  $ \pm 1332\ \mathrm{cm}^{-1}$. The error bars are given by $\sqrt{N}$, where $N$ is the total number of counts. The stars placed at $\delta\tau= 9\ \mathrm{ps}$ represent the individual contribution to the $I_{SaS}^{Real}$ intensity from the H-polarized (pink) and V-polarized (green) beams.\label{fig3}} 
\end{figure}

We measure the SaS scattering intensity  ($I_{SaS}$), which is the number of coincidence counts observed in the $\Delta\tau = 0$ peak in the histogram,  by varying the time delay $\delta\tau$ between the H-pulse and the V-pulse from $-2\ \mathrm{ps}$ to $+13\ \mathrm{ps}$. The results are shown in Figure \ref{fig3}. The red circles are the intensities of the \textit{real} SaS ($I_{SaS}^{Real}$) and the blue circles are the results for the \textit{virtual} SaS ($I_{SaS}^{Virtual}$) process.     

We started the experiments overlapping the pulses H and V  (delay time $\delta\tau=0$) by searching for the maximum luminescence of a graphene sample~\cite{Lui2010}. We then changed the sample to diamond and varied the length of the delay line, thus varying the delay time between the pulses. For negative delay time values, the V pulse reaches the sample first, followed by the H pulse. 

Analyzing the \textit{real} SaS (red circles in Fig. \ref{fig3}), the intensity $I_{SaS}^{Real}$ starts to increase when the pulses start to overlap, reaches a maximum for a delay time $\delta\tau\approx 0.5\ \mathrm{ps}$, and decreases for larger delays. The constant non-zero $I_{SaS}^{Real}\approx 0.2$ value is related to SaS processes from each pulse individually and amounts to the much less probable but still existing cases when either H or V-polarized pulses generates SaS pairs with crossed polarizations. To test this hypothesis, we have measured the intensities of $(H;H,V)$ and $(V;H,V)$ at $\delta\tau=9\ \mathrm{ps}$. The results are shown in Fig. \ref{fig3} by the star-symbols at $\delta\tau=9\ \mathrm{ps}$ and, summed up, account for the $I_{SaS}^{Real}\approx 0.2$ background. 

The $\delta\tau$ dependence of the real SaS process can be explained considering the sample excited by a first laser pulse with an intensity Gaussian profile $I(t) \propto \exp({- t^{2} /\sigma_L^{2}})$, where $\sigma_L=w_L/\sqrt{\ln 16}=(0.24\pm 0.01)\ \mathrm{ps}$. After the H-pulse reaches the sample, creating an S photon and a phonon, the probability of a V-pulse to annihilate this  phonon at a time delay $\delta\tau$, creating an aS photon is given by
\begin{equation}
\begin{split}
P_{SaS}\left(\delta\tau\right) &= A \sqrt{\frac{\pi}{2}}\, \sigma_L \int_{0}^{\infty} e^{-\left(t^{\prime}-\delta\tau \right)^2 / 2\sigma_L^{2}} e^{-t^{\prime} / \tau_{P}} d t^{\prime}\\
&=Be^{-\delta\tau/\tau_{P}}\left[1+\mathrm{erf}\!\left(\frac{\delta\tau}{\sqrt{2}\,\sigma_L}-\frac{\sigma_L}{\sqrt{2}\,\tau_{P}}\right)\right],
\end{split}
\label{eq1}
\end{equation}    
where $B= A\pi \sigma_L^{2} e^{\sigma_L^{2} /\left(2 \tau_{P}^{2}\right)} /2$, with $A$ depending on the Raman cross-sections, and $\mathrm{erf}$ is the error function. The SaS intensity as a function of delay time between laser pulses is given by 
\begin{equation}
I_{SaS}(\delta\tau) = C_{SaS} P_{SaS}(\delta\tau) + I^{SP}_{SaS},
\label{eq2}
\end{equation} 
where $I^{SP}_{SaS}$ is a constant that does not depend on $\delta\tau$ and represents the intensity of the SaS processes generated in a single pulse (stars in Fig. \ref{fig3}). In Fig. \ref{fig3}, the normalized intensity of \textit{real} SaS (red circles) is fitted by Eq. \ref{eq2}  (solid red line), with $\tau_{P}=(2.8 \pm 0.3)\ \mathrm{ps}$ and $I^{SP}_{SaS}=0.22 \pm 0.02$ as fitting parameters, for a laser pulse width of $0.40\ \mathrm{ps}$. Qualitatively, the process is maximized when there is enough time for the first pulse to create the phonon (hence the delay) and not enough time for this phonon to decay.

Considering now the \textit{virtual} SaS  (blue circles in Figure \ref{fig3}), the process is expected to happen within the overlap time of the H and V laser pulses.  By analyzing the blue data, we observe that, indeed, the intensity of the \textit{virtual} SaS is better fitted by the convolution of two Gaussians with time width of $0.40\ \mathrm{ps}$  FWHM (same time width of the laser pulse). When we increase or decrease $\delta\tau$, reducing the overlap of the pulses, the intensity of the SaS process decreases (solid blue line in Fig. 5).

\section{Conclsion}
\label{s:conc} 

In summary, based on the strong polarization correlation between the excitation laser and the scattered  SaS photon pairs,  we have elucidated a fundamental difference between \textit{real} and \textit{virtual} SaS pair production processes in a diamond sample. By means of a pump-and-probe experiment with cross-polarized and time-delayed laser pulses, we showed that the production rate of \textit{real} SaS pairs decreases with the decay of the phonon population generated by the Stokes process. We measure an SaS time correlation profile compatible with a phonon population lifetime $\tau_{P} = (2.8\pm 0.3)\ \mathrm{ps}$ in diamond \cite{Lee2011e, Lee2010}. Quite differently from the \textit{real} process, the \textit{virtual} SaS pair production occurs just as long as the pump and probe laser pulses overlap, indicating that it is faster than the duration of a single pulse, which is $0.40\ \mathrm{ps}$ FWHM in our experiment. Assuming that the virtual SaS pair production is a coherent process, the time correlation between S and aS photons in this regime is probably limited by the bandwidth of the detection system. Further investigation is needed to clarify this point. Furthermore, it is important to expand these findings to other materials/media for testing the generality of these results. Similar polarization behavior has been observed for water \cite{Kasperczyk2016}. However, the lifetime experiment in liquids is more challenging because of the smaller phonon lifetimes [REF] and the smaller SaS production rate \cite{Junior2019}.

This work was supported by CNPq (305384/2015-5, 429165/2018-8, 302775/2018-8, INCT-IQ 465469/2014-0) and FAPERJ Project E-26/202.290/2018.

\nocite{*}

%\bibliographystyle{ieeetr}
%\bibliography{lifetime}% Produces the bibliography via BibTeX.

\end{document}